\documentclass[a4paper,11pt]{article}
\usepackage{jheppub} 
\usepackage{lineno}
\usepackage{amsmath}
\usepackage{physics}
\usepackage{subfigure}
\usepackage{mathrsfs}
\usepackage{placeins}
\usepackage[normalem]{ulem}
\usepackage{xcolor}

\def\bc{\begin{center}}

\def\ec{\end{center}}
\def\be{\begin{eqnarray}}
\def\ee{\end{eqnarray}}

\def\blue{\color{blue}}

\newcommand{\Ket}[1]{\left |#1\right )}
\newcommand{\Bra}[1]{\left (#1\right |}

\title{\boldmath Krylov complexity and Wightman power spectrum with positive chemical potential in Schr\"odinger field theory}

\author[a]{Peng-Zhang He}
\author[b]{,Lei-Hua Liu}
\author[c,d,1]{,Hai-Qing Zhang\note{Corresponding author.}}
\author[a,2]{,Qing-Quan Jiang\note{Corresponding author.}}
\affiliation[a]{School of Physics and Astronomy, China West Normal University, Nanchong 637002, Sichuan, China}
\affiliation[b]{Department of Physics, College of Physics, Mechanical and Electrical Engineering, Jishou University, Jishou 416000, China}
\affiliation[c]{Center for Gravitational Physics, Department of Space Science, Beihang University, Beijing 100191, China}
\affiliation[d]{Peng Huanwu Collaborative Center for Research and Education, Beihang University, Beijing 100191, China}

\emailAdd{hepzh@cwnu.edu.cn}
\emailAdd{liuleihua8899@hotmail.com}
\emailAdd{hqzhang@buaa.edu.cn}
\emailAdd{qqjiangphys@yeah.net}

\abstract{We study Krylov complexity in Schr\"odinger field theory in the grand canonical ensemble with chemical potential $\mu$, with an emphasis on the qualitatively new features that arise for $\mu>0$. In this regime the fermionic Wightman power spectrum is effectively single-sided and sharply truncated at $\omega=\mu$, which induces a crossover in the Lanczos coefficients {and signals a dynamical transition from a bulk-dominated regime to a spectral-edge-dominated regime}: $b_n$ displays a two-stage linear growth (from an early-time slope $\pi/\beta$ to an asymptotic slope $2/\beta$), while $a_n$ bends from near-zero values to a linear descent with slope $-4/\beta$. We provide analytic support for the resulting complexity growth from three complementary viewpoints: (i) using an $SL(2,\mathbb{R})$ algebraic construction matched to the asymptotic Lanczos data, we show that the late-time Krylov complexity must grow quadratically, $K(t)\propto t^{2}$; (ii) by analyzing engineered Wightman spectra with controlled decay and truncation, we identify single-sided exponential decay as the key spectral feature responsible for the quadratic asymptotics, while an approximately even two-sided exponential spectrum explains the early-time $K(t)\sim\sinh^{2}(\pi t/\beta)$ behavior at large $\mu$; (iii) we formulate the problem in terms of orthogonal polynomials and estimate the crossover scale separating the early- and late-stage regimes. Overall, our results help clarify the role of chemical potential and spectral truncation in shaping operator growth and Krylov complexity in this non-relativistic quantum field theory setting.}

\begin{document}
\maketitle
\flushbottom
   \section{Introduction}
   Krylov complexity is a quantitative probe of operator growth in quantum many-body systems. It provides a complementary viewpoint on information spreading, quantum chaos and thermalization, and has found applications ranging from field theory to cosmology and non-Hermitian and $T\bar T$-deformed systems (see e.g. \cite{Parker:2018yvk,Dymarsky:2021bjq,Trigueros:2021rwj,Rabinovici:2020ryf,Rabinovici:2021qqt,Rabinovici:2022beu,Bhattacharjee:2022vlt,Khetrapal:2022dzy,Liu:2022god,Avdoshkin:2022xuw,Fan:2023ohh,Mohan:2023btr,Chen:2024imd,Baggioli:2024wbz,Alishahiha:2024rwm,Qi:2024tau,Camargo:2024deu,Aguilar-Gutierrez:2025hbf,Camargo:2024rrj,Adhikari:2022oxr,Li:2024kfm,Li:2024iji,Li:2024ljz,Zhai:2024tkz,Zhai:2025abc,Erdmenger:2023wjg,Nandy:2024mml,He:2022ryk,Chattopadhyay:2024pdj,Bhattacharya:2022gbz,Bhattacharjee:2022ave,Bhattacharjee:2022lzy,Bhattacharjee:2023uwx,Bhattacharya:2023zqt,Craps:2024suj,Malvimat:2024vhr,Vasli:2023syq,Kundu:2023hbk,Adhikari:2022whf,Lv:2023jbv,Xu:2024gfm,Camargo:2023eev,Hornedal:2022pkc,Caputa:2021sib,Caputa:2024vrn,Bento:2023bjn,Dymarsky:2019elm}. Interested readers can also consult recent reviews \cite{Nandy:2024evd,Rabinovici:2025otw}.)

   The Krylov-space approach is based on the Lanczos algorithm \cite{viswanath1994recursion}, which recursively generates a Krylov basis and the associated Lanczos coefficients $a_n$ and $b_n$ encoding how an operator spreads along the Krylov chain. A central expectation in thermal quantum systems is the ``universal operator growth'' picture: in many chaotic systems $b_n$ grows linearly at large $n$, which in turn leads to an exponential growth of Krylov complexity $K(t)$ \cite{Parker:2018yvk}; in quantum field theory, a thermal universal slope $b_n\sim \pi n/\beta$ was established in \cite{Dymarsky:2021bjq}, again implying exponential growth (see also \cite{Camargo:2022rnt,He:2024xjp}).

   At the forefront of quantum field theory and quantum gravity, the connection between Krylov complexity and holographic duality \cite{Maldacena:1997re} has become a hotspot of research recently. In the framework of AdS/CFT duality, the correspondence between the double-scaled SYK (DSSYK) model and Jackiw-Teitelboim (JT) gravity in the bulk has provided an explicit holographic example for the Krylov complexity in \cite{Rabinovici:2023yex}. It was proposed that the Krylov complexity of the infinite-temperature thermofield double state (TFD) on the AdS$_{2}$ boundary precisely corresponds to the geometric length of the double-sided wormhole in the bulk gravitational theory. The Krylov complexity is also named as spread complexity in the state formalism \cite{Balasubramanian:2022tpr,Caputa:2022yju,Huh:2023jxt,Ganguli:2024uiq,Fu:2024fdm}.	Recent studies suggest that the time derivative of spread complexity could be proportional to the proper momentum of particles in AdS spacetime \cite{Caputa:2024sux,Fan:2024iop,He:2024pox,Caputa:2025dep}. The authors of \cite{Caputa:2024sux} proposed this conjecture and demonstrated its validity for a massive particle in three-dimensional AdS spacetime. Soon thereafter, this relationship was also confirmed in higher-dimensional AdS spacetime, and regardless of whether the particles are massive or not, the growth rate of the spread complexity of the boundary CFT is proportional to the radial momentum of the particles \cite{Fan:2024iop,He:2024pox}. Furthermore, the authors of \cite{Caputa:2025dep} revisit local quench dynamics in two-dimensional conformal field theories using the Krylov space approach and AdS/BCFT correspondence. They demonstrate that the tip of the end-of-the-world (EOW) brane in the bulk—both in vacuum and at finite temperature or finite size—moves radially according to the equation of motion for a massive particle in the Ba\~nados geometry. If a particle with mass $m=c/16$ is associated with the tip of the EOW brane, its proper radial momentum is also proportional to the growth rate of spread complexity in the boundary CFT. Very recently, building on the pioneering work of \cite{Caputa:2024sux}, literature \cite{Li:2025fqz} utilized $SL(2,\mathbb{R})$ symmetry to construct Krylov basis and demonstrated that their non-uniqueness leads to an ambiguity analogous to quantum complexity. This work further proposed a more fundamental and universal holographic interpretation for spread complexity and its rate of growth in two-dimensional conformal field theories: spread complexity is interpreted as the particle energy measured by a bulk observer, while its rate of growth corresponds to the radial momentum.

   {\noindent\textbf{Schr\"odinger field theory.} Krylov complexity in the grand canonical ensemble was previously studied in the regime of non-positive chemical potential ($\mu\le 0$) \cite{He:2024hkw}. In that earlier work the late-time behavior was interpreted as an exponential growth; in the present paper we clarify that the correct asymptotics is instead \emph{quadratic}. This clarification is supported from two complementary viewpoints developed here: the $SL(2,\mathbb{R})$ analysis based on the large-$n$ slopes of Lanczos coefficients (Sec.~\ref{sec3}) and the explicit solvable example of a single-sided exponentially decaying Wightman spectrum leading to $K(t)\propto t^{2}$ (Sec.~\ref{sec4}, Subsec.~\ref{subsec:single_sided_exp}). For completeness, we summarize/revisit the $\mu\le 0$ setup and results in Sec.~\ref{sec2}.

   The \emph{positive} chemical potential regime, by contrast, is qualitatively different and largely unexplored: for $\mu>0$ the fermionic Wightman power spectrum becomes effectively single-sided and sharply truncated at the spectral edge $\omega=\mu$, and we show that this hard edge induces a dynamical transition in operator growth, manifested as a two-stage linear growth of $b_n$ and a deflection of $a_n$. Similar ``turning point'' phenomena triggered by spectral truncation were emphasized recently in Refs.~\cite{Teretenkov:2024uwm,Uskov:2024xac}, where the mechanism controlling the emergence/location of the turning point was identified as an open problem.

   \noindent\textbf{This paper.} We investigate Krylov complexity for fermionic fields in Schr\"odinger field theory at \emph{positive} chemical potential $\mu$, and use engineered Wightman power spectra and orthogonal-polynomial methods to isolate the spectral mechanism behind the observed ``turning point''/deflection in the Lanczos coefficients \cite{Teretenkov:2024uwm,Uskov:2024xac}. Our main findings are:
   \begin{itemize}
   \item \textbf{Numerical Lanczos data and crossover.} For $\mu>0$, the Wightman spectrum is effectively single-sided and sharply truncated at $\omega=\mu$. Correspondingly, $a_n$ bends from near-zero values to a linear decrease with slope $-4/\beta$, while $b_n$ shows a two-stage linear growth, changing from an early-time slope $\pi/\beta$ to an asymptotic slope $2/\beta$. The deflection point moves to larger $n$ as $\mu$ increases.
   \item \textbf{Late-time complexity from asymptotic Lanczos coefficients.} Matching the large-$n$ Lanczos behavior to an $SL(2,\mathbb{R})$ algebraic construction, we show that the late-time growth becomes quadratic, $K(t)\propto t^{2}$. We emphasize that this is consistent with linear $b_n$ because the large-$n$ slopes satisfy the degenerate condition $\gamma^{2}=4\alpha^{2}$, which collapses the would-be exponential channel to a polynomial one.
   \item \textbf{Spectral mechanism and universality.} Using engineered Wightman spectra with controlled decay and truncation, we identify single-sided exponential decay as the key spectral feature behind the quadratic asymptotics, while an approximately even two-sided exponential spectrum explains the early-time $K(t)\sim\sinh^{2}(\pi t/\beta)$ behavior at large $\mu$.
   \item \textbf{Crossover estimate from orthogonal polynomials.} Formulating the problem in terms of orthogonal polynomials with the Wightman spectrum as the weight function, and combining this with the {\it Gershgorin disk theorem}, we estimate the crossover scale separating the bulk-dominated and spectral-edge-dominated regimes and relate it quantitatively to the spectral edge at $\omega=\mu$.
   \end{itemize}

   \noindent\textbf{Organization.} Sec.~\ref{sec2} reviews the Krylov-space framework and the Schr\"odinger field theory setup; in particular, the discussion of the $\mu\le 0$ regime is included mainly for completeness and closely follows Ref.~\cite{He:2024hkw}. Sec.~\ref{sec3} presents the new numerical Lanczos data and Krylov-complexity evolution for {positive} chemical potential (with moments evaluated using the series truncation $\sum_{k=0}^{200}(\cdots)$). Sec.~\ref{sec4} analyzes engineered Wightman spectra to isolate the spectral mechanism behind deflection/plateau behaviors and the resulting complexity growth. Sec.~\ref{sec5} draws the conclusions. Appendix~\ref{app:crossover} provides an orthogonal-polynomial derivation and crossover estimate; Appendix~\ref{appa} discusses the sign flip of $a_n$ under $f^{W}(\omega)\rightarrow f^{W}(-\omega)$; Appendix~\ref{appb} studies double-inflection behavior under asymmetric truncation.}

   \section{Krylov complexity of the Schr\"odinger field theory with non-positive chemical potential}\label{sec2}
   This section will provide a brief introduction to the Krylov complexity, and then discuss the Krylov complexity of the Schr\"odinger field theory with non-positive chemical potential. {Much of the material here is included for completeness and closely follows our earlier work~\cite{He:2024hkw}.}
   
   \subsection{Lanczos algorithm and Krylov complexity}\label{sec2.1}
   We start with a brief introduction to the Lanczos algorithm and its connection to Krylov complexity. The latter is a measure that characterizes how quantum operators grow as time evolves. Let's consider a unitary evolution of an operator $\mathcal{O}(t)$ under a time-independent Hamiltonian $H$,
   \begin{equation}\label{2.111}
   	\mathcal{O}(t)=e^{iHt}\mathcal{O}(0)e^{-iHt}\equiv e^{i\mathcal{L}t}\mathcal{O}(0),\qquad \mathcal{L}:=[H,\cdot],
   \end{equation}
   where $\mathcal{L}$ is called the Liouvillian superoperator. Expand the equation \eqref{2.111} at $t = 0$, then we have
   \begin{equation}
   	\mathcal{O}(t)=\sum_{n=0}^{\infty}\frac{(it)^n}{n!}\mathcal{L}^n\mathcal{O}(0).
   \end{equation}
   The above equation can be regarded as the expansion of the operator $\mathcal{O}(t)$ in a basis set $\{\mathcal{L}^n\mathcal{O}(0)\}$. The space spanned by this basis set is called the Krylov space. However, in general the set $\{\mathcal{L}^n\mathcal{O}(0)\}$ does not form an orthogonal basis. Therefore, one can apply the Gram-Schmidt procedure to obtain an orthonormal basis, namely the Krylov basis $\{\mathcal{O}_n\}$ \cite{Parker:2018yvk,viswanath1994recursion}. The procedure is as follows:
   \begin{gather}
   	\mathcal{L}\Ket{\mathcal{O}_0}=a_{0}\Ket{\mathcal{O}_0}+b_{1}\Ket{\mathcal{O}_{1}},\\
   	\mathcal{L}\Ket{\mathcal{O}_{1}}=b_{1}\Ket{\mathcal{O}_{0}}+a_{1}\Ket{\mathcal{O}_{1}}+b_{2}\Ket{\mathcal{O}_{2}},\\
   	\cdots\notag\\
   	\mathcal{L}\Ket{\mathcal{O}_n}=b_{n}\Ket{\mathcal{O}_{n-1}}+a_{n}\Ket{\mathcal{O}_{n}}+b_{n+1}\Ket{\mathcal{O}_{n+1}},\label{2.6}\\
   	\cdots\notag
   \end{gather}
    Here, we use $\Ket{\mathcal{O}_n}$ to represent the Krylov basis $\mathcal{O}_n$, and take $\mathcal{O}_0 \equiv \mathcal{O}(0)$. Moreover, $\{a_n\}$ and $\{b_n\}$ are referred to as the Lanczos coefficients, which are defined as
    \begin{equation}
    	a_{n}=\Bra{\mathcal{O}_n}\mathcal{L}\Ket{\mathcal{O}_{n}},\qquad b_{n}=\Bra{\mathcal{O}_{n}}\mathcal{L}\Ket{\mathcal{O}_{n-1}},\qquad b_{0}\equiv0,
    \end{equation}
    where the inner product $\left(\cdot|\cdot\right)$ represents the Wightman inner product,
    \begin{equation}\label{2.7}
    	\left(A|B\right)=\expval{e^{H\beta /2}A^{\dagger}e^{-H\beta/2}B}_{\beta}\equiv\text{Tr}\left (e^{-\beta H}e^{H\beta /2}A^{\dagger}e^{-H\beta/2}B\right ),
    \end{equation}
     where $\beta$ is the inverse temperature. It follows from Eq.~\eqref{2.7} that the Liouville superoperator is Hermitian in Krylov space. The operator $\Ket{\mathcal{O}(t)}$ can be expanded in the Krylov basis as
    \begin{equation}
    	\Ket{\mathcal{O}(t)}=\sum_{n=0}^{\infty}i^n\varphi_{n}(t)\Ket{\mathcal{O}_{n}},\qquad {\rm with}~~\sum_{n=0}^{\infty}\abs{\varphi_{n}(t)}^{2}=1.
    \end{equation}
   Combining the Heisenberg equation
   \begin{equation}
   	\frac{d}{dt}\mathcal{O}(t)=i\mathcal{L}\mathcal{O}(t)
   \end{equation}
   with the recursion relations \eqref{2.6}, one obtains the time evolution of $\varphi_n$,
    \begin{equation}\label{2.12}
    	\partial_t\varphi_n(t)=ia_n\varphi_n(t)+b_{n}\varphi_{n-1}(t)-b_{n+1}\varphi_{n+1}(t).
    \end{equation}
  This is the discrete Schr\"odinger equation, showing that operator growth in Krylov space can be viewed as a hopping problem on a one-dimensional chain, with $n$ playing the role of the lattice site. The operator that measures Krylov complexity is defined as
    \begin{equation}
    	\hat{\mathcal{K}}=\sum_{n}n\Ket{\mathcal{O}_n}\Bra{\mathcal{O}_n}.
    \end{equation}
    The Krylov complexity is then defined by
    \begin{equation}\label{def:KrylovComplexity}
    	K(t)=\Bra{\mathcal{O}(t)}\hat{\mathcal{K}}\Ket{\mathcal{O}(t)}=\sum_{n}n\abs{\varphi_{n}(t)}^{2}.
    \end{equation}
 From this definition, $K(t)$ is the average position of the wave function $\varphi_n$ on the Krylov chain. {In all plots of Krylov complexity in this paper, we show $1+K(t)$ (instead of $K(t)$) for numerical convenience.}

   To compute the Krylov complexity, it is necessary to first determine the Lanczos coefficients and solve the discrete Schr\"odinger equation \eqref{2.12}. The Lanczos coefficients can be computed by adopting the moment method with the moments defined as
   \begin{equation}\label{2.16}
   	\mu_{n}=\Bra{\mathcal{O}(0)}\mathcal{L}^{n}\Ket{\mathcal{O}(0)}=\Bra{\mathcal{O}(0)}(-i)^{n}\dfrac{d^{n}}{dt^{n}}\Ket{\mathcal{O}(t)}|_{t=0}=(-i)^{n}\frac{d^{n}}{dt^{n}}\varphi_{0}(t)|_{t=0}.
   \end{equation}
   Since $\mathcal{L}$ is Hermitian in Krylov space, the moments are real. To compute them, we introduce the autocorrelation function
   \begin{equation}
   	C(t):=\left(\mathcal{O}(t)|\mathcal{O}(0)\right)=\varphi_{0}^{\ast}(t)
   \end{equation}
   and define the Wightman power spectrum as its Fourier transform,
   \begin{equation}
   	f^{W}(\omega)=\int_{-\infty}^\infty dt \, e^{i\omega t}C(t).
   \end{equation}
   The moments \eqref{2.16} are then related to the power spectrum by
   \begin{equation}\label{moments}
   	\mu_{n}=\frac{1}{2\pi}\int_{-\infty}^{\infty}f^{W}(\omega)\omega^{n}d\omega.
   \end{equation}
   Noting that $\mu_0 = 1$ and the power spectrum satisfies the normalization condition
   \begin{equation}\label{2.188}
   	1=\frac{1}{2\pi}\int_{-\infty}^{\infty}f^{W}(\omega)d\omega.
   \end{equation}
   As long as we have the moments, we can use the moment method to obtain the Lanczos coefficients. The moment method explicitly establishes the following nonlinear recurrence relations between the moments and the Lanczos coefficients,
   \begin{eqnarray}
   	M^{(n)}_{k} &=& L^{(n-1)}_{k} - L^{(n-1)}_{n-1} \frac{M^{(n-1)}_{k}}{M^{(n-1)}_{n-1}}, \\
   	L^{(n)}_{k} &=& \frac{M^{(n)}_{k+1}}{M^{(n)}_{n}} - \frac{M^{(n-1)}_{k}}{M^{(n-1)}_{n-1}},
   \end{eqnarray}
   where \( n = 1, \cdots, 2K \), \( k = n, \cdots, 2K - n + 1 \), where $K$ is a large integer. The initial conditions are given by
   \begin{equation}
   	M^{(0)}_{k} = (-1)^{k} \mu_{k}, \qquad L^{(0)}_{k} = (-1)^{k+1} \mu_{k+1}, \qquad k = 0, \cdots, 2K.
   \end{equation}
   From these relations, the resulting Lanczos coefficients are obtained as
   \begin{equation}
   	b_{n}^{2} = M^{(n)}_{n}, \qquad a_{n} = -L^{(n)}_{n}, \qquad n = 0, \cdots, K.
   \end{equation}
   These seemingly complex relations can be easily computed by using Mathematica.

   \subsection{Krylov complexity of the Schr\"odinger field theory with non-positive chemical potential}\label{sec2-2}
   For the Schr\"odinger field theory, the Lagrangian is given by \cite{altland2010condensed}:
   \begin{equation}
   	\mathscr{L}=\psi^{\dagger}\left (i\frac{\partial}{\partial t}+\frac{\nabla^{2}}{2m}\right )\psi,
   \end{equation}
   where $m$ is the mass of non-relativistic free boson or fermion\footnote{If $\psi$ satisfies canonical commutation relations, it describes identical non-relativistic bosons; conversely, if $\psi$ obeys canonical anti-commutation relations, it represents identical fermions \cite{Mintchev:2022xqh}.}. This Lagrangian exhibits symmetry under the global $U(1)$ transformation
   \begin{equation}
   	\psi \rightarrow e^{i\alpha} \psi, \qquad \psi^{\dagger} \rightarrow \psi^{\dagger} e^{-i\alpha}.
   \end{equation}
   The conserved charge associated with this symmetry is
   \begin{equation}
   	N = \int d^{d-1} \mathbf{x} \, \mathscr{N} = \int d^{d-1} \mathbf{x} \, \psi^{\dagger} \psi,
   \end{equation}
   where $\mathscr{N}$ denotes the charge density. Therefore, we can consider the Schr\"odinger field theory in the grand canonical ensemble. By replacing the Hamiltonian as $H \rightarrow H - \mu N$ in Sec.~\ref{sec2.1},\footnote{Please do not confuse the chemical potential $\mu$ here with the moments $\mu_n$ defined in Eq.~\eqref{2.16}.} we obtain the corresponding Lanczos coefficients and Krylov basis in the grand canonical ensemble. The Wightman power spectrum of the Schr\"odinger field theory becomes \cite{He:2024hkw}
   \begin{equation}
   	f^{W}(\omega)=\mathcal{N}(\mu-\omega)^{\frac{d-3}{2}}\frac{\eta  e^{-\frac{\beta  \omega }{2}}}{\eta  e^{-\beta  \omega }-1}\Theta(\mu-\omega),
   \end{equation}
   where $\eta=1$ for bosons and $\eta=-1$ for fermions, and $\mathcal{N}$ is the normalization constant determined by the normalization condition \eqref{2.188}. In the rest of this paper, we focus on $d=5$, which is simple and representative (other dimensions can be treated similarly). In this case, the power spectrum is
   	\begin{equation}\label{2.27}
   		f^{W}(\omega)=\left \{\begin{matrix}
   			\mathcal{N}(\mu-\omega)e^{-\frac{\beta\omega}{2}}n_{B}(\omega)\Theta(\mu-\omega),&\qquad&\text{Bosonic field},\\
   			\mathcal{N}(\mu-\omega)e^{-\frac{\beta \omega}{2}}n_{F}(\omega)\Theta(\mu-\omega),&\qquad &\text{Fermionic field},
   		\end{matrix}\right .
   	\end{equation}
   where $n_{B}(\omega)=1/(e^{\beta\omega}-1)$ is the Bose–Einstein distribution and $n_{F}(\omega)=1/(e^{\beta\omega}+1)$ is the Fermi–Dirac distribution \cite{Kapusta:2007xjq}. For the Bose-Einstein distribution, the chemical potential can only be less than zero. For the Fermi-Dirac distribution, the chemical potential can be positive, negative, or zero. 
   
   The Krylov complexity of the Schr\"odinger field theory with non-positive chemical potential was studied in Ref.~\cite{He:2024hkw}. It was found that for non-positive chemical potentials, for both bosonic and fermionic fields, the Lanczos coefficients exhibit a linear behavior without staggering and can be {approximated} by
   \begin{eqnarray}
   	\beta a_{n}&\approx&-4(n+1)+\mu,\\
   	\beta b_{n}&\approx& 2n+1.
   \end{eqnarray}
 {The corresponding Krylov complexity exhibits an asymptotic \emph{quadratic} growth at late times.} {This behavior should be understood as a consequence of the simultaneous large-$n$ scalings of $a_n$ and $b_n$ (rather than of $b_n$ alone): in the effective $SL(2,\mathbb{R})$ description, the special relation $\gamma^{2}=4\alpha^{2}$ leads to a degenerate limit in which the exponential-growth channel disappears and the late-time growth becomes quadratic.} Moreover, across different values of the chemical potential and for both bosonic and fermionic fields, the complexities show very similar behavior. In this paper, we explain why the complexity exhibits this similar behavior for negative chemical potential. {The theoretical support of this statement can be traced to: (i) the $SL(2,\mathbb{R})$ asymptotic analysis in Sec.~\ref{sec3} based on the large-$n$ slopes of the Lanczos coefficients (see around Eqs.~\eqref{2.18}--\eqref{2.19} and the condition $\gamma^{2}=4\alpha^{2}$), and (ii) the explicit example of a single-sided exponentially decaying spectrum in Sec.~\ref{sec4} (\emph{the first subsection, ``The single-sided exponentially decaying Wightman power spectrum''}), which yields $K(t)\propto t^{2}$.}

   \section{The Krylov complexity of the fermionic field with positive chemical potential}\label{sec3}
 {In this section we present the main new numerical results of this paper, namely the Krylov complexity and Lanczos data for the fermionic field at} {positive chemical potential} {in the Schr\"odinger field theory.} The Wightman power spectrum \eqref{2.27} of the fermionic field can be written as \footnote{Here we only consider the fermionic field rather than the bosonic field since the chemical potential in the Bose-Einstein distribution cannot be positive \cite{landau2013statistical}.} 
   \begin{equation}\label{2.1}
   \begin{aligned}
   		f^W(\omega)&\equiv g(\omega)\Theta(\mu-\omega)\\
   		&=\mathcal{N}e^{-\beta \omega/2}n_F(\omega)(\mu-\omega)\Theta(\mu-\omega)\\
   		&=\mathcal{N}\frac{e^{\beta\omega/2}}{1+e^{\beta\omega}}(\mu-\omega)\Theta(\mu-\omega),
   \end{aligned}
   \end{equation}
   where $g(\omega)=\mathcal{N}e^{-\beta \omega/2}n_F(\omega)(\mu-\omega)$ and the chemical potential $\mu$ is now positive. The normalization condition \eqref{2.188} becomes
   \begin{equation}\label{2.2}
   	\begin{aligned}
   		1&=\frac{1}{2\pi}\int_{-\infty}^{\infty}f^{W}(\omega)d\omega\\
   		&=\frac{1}{2\pi}\int_{-\infty}^{\infty}g(\omega)d\omega-\frac{1}{2\pi}\int_{\mu}^{\infty}g(\omega)d\omega\\
   		&=\frac{1}{2\pi}\int_{-\infty}^{\infty}g(\omega)d\omega-\frac{1}{2\pi}\int_{\mu}^{\infty}\mathcal{N}(\mu-\omega)\sum_{k=0}^{\infty}(-1)^{k}e^{-\beta \omega(k+1/2)}d\omega.
   	\end{aligned}
   \end{equation}
  The above condition can be used to determine the normalization constant $\mathcal{N}$. In the numerical calculations, we truncate the series to obtain an accurate approximation. In practice, we set the upper limit of the summation to $k=200$ {(i.e. we sum over $k=0,1,\ldots,200$)}, which is sufficient for our purposes. Similarly, the moments in \eqref{moments} can be computed using the same truncation,
   \begin{equation}\label{2.3}
   	\mu_{n}=\frac{1}{2\pi}\int_{-\infty}^{\infty}g(\omega)\omega^{n}d\omega-\frac{1}{2\pi}\int_{\mu}^{\infty}\mathcal{N}_{200}(\mu-\omega)\omega^{n}\sum_{k=0}^{200}(-1)^{k}e^{-\beta\omega(k+1/2)}d\omega,
   \end{equation}
   where $\mathcal{N}_{200}$ is referred to as the normalization constant with the upper limit $k=200$ in \eqref{2.2}. It is worth noting that when $\mu$ is sufficiently large, the power spectrum is given by
   \begin{equation}\label{2.4}
   	f^{W}(\omega)=\mathcal{N}\frac{e^{\beta\omega/2}}{1+e^{\beta\omega}} 
   \end{equation}
   for $\abs{\omega} \ll \mu$. \footnote{Note that we have absorbed $\mu$ into the normalization constant $\mathcal{N}$ in equation \eqref{2.4}.} That is to say, when $\mu$ is sufficiently large, the power spectrum behaves as an even function  in the low frequency limit $\abs{\omega}\ll\mu$. Therefore, {\textit{as $\mu\rightarrow\infty$, the power spectrum will also exhibit the behavior as an even function}}, which is \eqref{2.4}. We will discuss the relation between the Wightman power spectrum and the Krylov complexity in depth in the section \ref{sec4}. 
   
   \subsection{Lanczos coefficients}
   
\begin{figure}[htb]
   	\centering
   	\subfigure[Lanczos coefficient $a_n$.]{\includegraphics[width=0.45\linewidth]{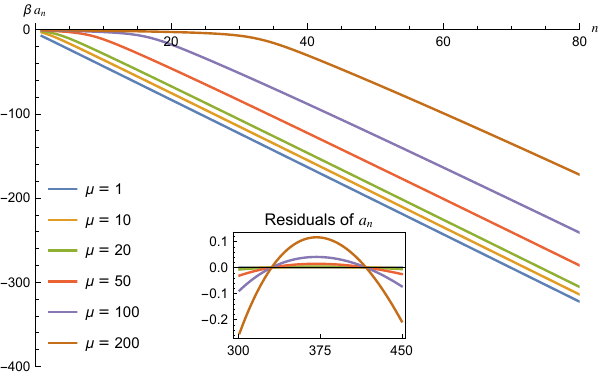}
   	\label{fig:anlist}}
   	\hfill
   	\subfigure[Lanczos coefficient $b_n$.]{\includegraphics[width=0.45\linewidth]{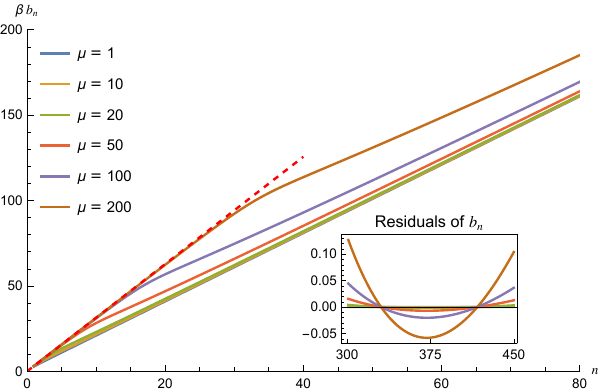}
	\label{fig:bnlist}}
	\subfigure[The slopes of $\beta a_n$ for various chemical potentials.]{\includegraphics[width=0.45\linewidth]{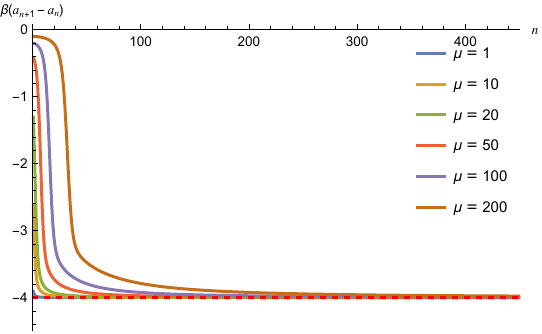}
	\label{fig:danlist}}
   	\hfill
   	\subfigure[The slopes of $\beta b_n$ for various chemical potentials.]{\includegraphics[width=0.45\linewidth]{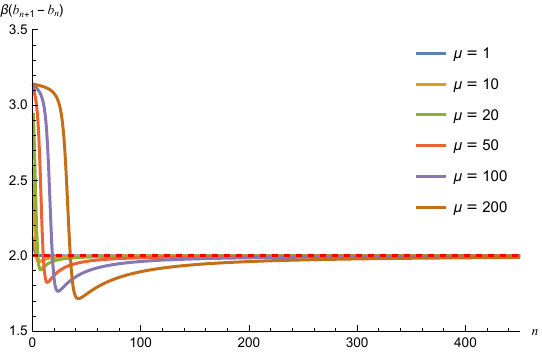}
	\label{fig:dbnlist}}
   		\caption{{The Lanczos coefficients $\{a_n\}$ (panel (a)) and $\{b_n\}$ (panel (b)) for chemical potentials $\mu = 1, 10, 20, 50, 100, 200$. All numerical moments entering the Lanczos construction are evaluated with the series truncation $\sum_{k=0}^{200}(\cdots)$ in Eq.~\eqref{2.3}. The red dashed line in panel (b) shows the early-stage linear growth $\beta b_n = \pi n$. The insets in panels (a) and (b) show the residuals of the corresponding linear fits, plotted using the numerical fit parameters reported in Table~\ref{tab:lanczos_fits_app}, {with the linear fits performed over the window $301\le n\le 450$}. Panels (c) and (d) display the (numerical) local slopes of $\beta a_n$ and $\beta b_n$ as functions of $n$; the red dashed lines at $-4$ and $2$ indicate the expected asymptotic constants.}}
   		\label{Lanczos}
   \end{figure}

   In the following, we will take the chemical potential $\mu=1, 10, 20, 50, 100, 200$ as examples to study the Lanczos coefficients of the fermionic field. Since the power spectrum \eqref{2.1} in general is not an even function, there are two sequences of the Lanczos coefficients, i.e. $a_n$ and $b_n$ \cite{He:2024hkw}. The numerical results of the Lanczos coefficients are shown in the Figure \ref{Lanczos}.
 From Figs.~\ref{fig:anlist} and \ref{fig:bnlist}, we find that the Lanczos coefficients $a_n$ and $b_n$ do not exhibit staggering, similar to the negative-$\mu$ results of Ref.~\cite{He:2024hkw}. However, the behavior for $\mu>0$ is qualitatively different. In particular, as shown in Fig.~\ref{fig:anlist}, $a_n$ stays close to zero over an initial range and then decreases linearly with $n$. The deflection point moves to larger $n$ as $\mu$ increases. By contrast, for negative chemical potentials, $a_n$ decreases linearly without any deflection (see Ref.~\cite{He:2024hkw}).
 
In Fig.~\ref{fig:bnlist}, we plot the Lanczos coefficient $b_n$ as a function of $n$ for various chemical potentials. The red dashed line corresponds to the early-stage behavior $\beta b_n = \pi n$. We see that $b_n$ first follows this line and then crosses over to a second linear regime with a different slope. The crossover point shifts to larger $n$ as $\mu$ increases. Moreover, the crossover in $b_n$ occurs at essentially the same $n$ where $a_n$ deflects. For example, for $\mu=200$, $b_n$ crosses over around $n\approx 32$, which is also the deflection point of $a_n$. {A quantitative estimate for the crossover location $n^{\ast}$ is given in Appendix~\ref{app:crossover}.} This behavior is markedly different from the negative-$\mu$ case, where $b_n$ for different chemical potentials collapse onto the same linear growth $\beta b_n=2n+1$ (see Ref.~\cite{He:2024hkw}). For $\mu>0$, by contrast, the $b_n$ curves neither collapse nor exhibit a single linear regime.

   
  We extract the local slopes of $\beta a_n$ and $\beta b_n$ from the discrete differences $\beta(a_{n+1}-a_n)$ and $\beta(b_{n+1}-b_n)$. The numerical results are shown in Figs.~\ref{fig:danlist} and \ref{fig:dbnlist}. We see that for all chemical potentials considered, the slopes approach the constants $-4$ and $2$, respectively, at sufficiently large $n$, consistent with the behaviors in Figs.~\ref{fig:anlist} and \ref{fig:bnlist}.

{This slope analysis implies the following universal large-$n$ asymptotic behaviors of the Lanczos coefficients.}
   \begin{equation}\label{3.5}
   	\beta a_n\sim -4 n+{\rm constant},~~~~~~n\gg1
   \end{equation}
   and
   \begin{equation}\label{3.6}
   	\beta b_n\sim 2 n+{\rm constant},~~~~~~n\gg1. 
   \end{equation}
{To quantitatively corroborate the asymptotic linear behaviors in Eqs.~\eqref{3.5} and \eqref{3.6} (and to facilitate later comparisons with the $SL(2,\mathbb{R})$ construction), we further perform linear regression analyses for both sequences $a_n$ and $b_n$ in the regime $301 \le n \le 450$. Concretely, we fit the numerical data to
\begin{equation}
	a_n \simeq s_1 n + c_1 \,,\quad b_n \simeq s_2 n + c_2 \,,
\end{equation}
with $s_{1,2}$ and $c_{1,2}$ denoting the slopes and intercepts, respectively. The fitted parameters, together with the standard errors (68\% confidence interval) and the adjusted coefficient of determination (Adjusted $R^2$), are summarized in Table~\ref{tab:lanczos_fits_app}. We find excellent linearity (Adjusted $R^2\simeq1$) and slopes consistent with the universal values $\beta s_1\simeq-4$ and $\beta s_2\simeq2$.

\begin{table}[ht]
	\centering
	\caption{Quantitative linear regression results for Lanczos coefficients in the asymptotic regime ($301 \le n \le 450$). The data are fitted to $a_n = s_1 n + c_1$ and $b_n = s_2 n + c_2$. The standard errors (SE) correspond to the slopes $s_1$ and $s_2$.}
	\label{tab:lanczos_fits_app}
	\renewcommand{\arraystretch}{1.2}
	\setlength{\tabcolsep}{5pt}
	\begin{tabular}{c c c c c c c c}
		\hline
		$\mu$ & $s_1$ ($a_n$) & $c_1$ ($a_n$) & SE($s_1$) & $s_2$ ($b_n$) & $c_2$ ($b_n$) & SE($s_2$) & Adj. $R^2$ \\
		\hline
		1   & $-4.0000$ & $-3.02$  & $1.4 \times 10^{-7}$ & $2.0000$ & $1.01$  & $5.4 \times 10^{-8}$ & $1.0000$ \\
		10  & $-3.9997$ & $5.63$   & $2.2 \times 10^{-6}$ & $1.9998$ & $1.19$  & $1.1 \times 10^{-6}$ & $1.0000$ \\
		20  & $-3.9991$ & $14.95$  & $6.1 \times 10^{-6}$ & $1.9995$ & $1.52$  & $3.0 \times 10^{-6}$ & $1.0000$ \\
		50  & $-3.9963$ & $41.84$  & $2.4 \times 10^{-5}$ & $1.9981$ & $3.08$  & $1.2 \times 10^{-5}$ & $1.0000$ \\
		100 & $-3.9894$ & $84.17$  & $6.9 \times 10^{-5}$ & $1.9947$ & $6.92$  & $3.5 \times 10^{-5}$ & $1.0000$ \\
		200 & $-3.9696$ & $162.30$ & $2.0 \times 10^{-4}$ & $1.9848$ & $17.87$ & $9.9 \times 10^{-5}$ & $1.0000$ \\
		\hline
	\end{tabular}
\end{table}
The results in Table~\ref{tab:lanczos_fits_app} indicate that the linear regressions provide an excellent description of the numerical Lanczos coefficients in the fitting window. To further assess the fit quality, we have added residual insets in Fig.~\ref{fig:anlist} and Fig.~\ref{fig:bnlist}. The residuals of $a_n$ exhibit an inverted ``U'' shape, while those of $b_n$ exhibit a ``U'' shape. We interpret these systematic curvatures as finite-$n$ effects originating from the low-frequency part of the spectrum, which makes it difficult for a finite-$n$ dataset to perfectly capture the strictly asymptotic linear regime.
}


   


   \subsection{Krylov complexity}\label{sec2.2}
\begin{figure}[htb]
   	\centering
   	\includegraphics[width=0.6\linewidth]{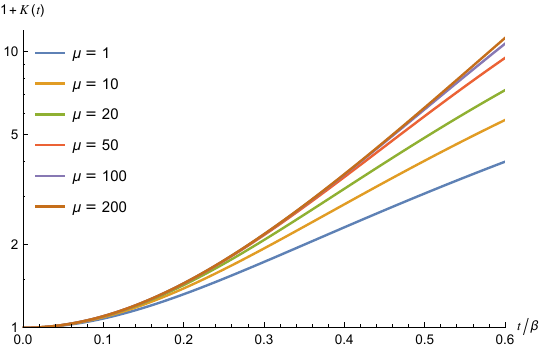}
   	\caption{Time evolution of the Krylov complexity with chemical potentials $\mu = 1, 10, 20, 50$, $100, 200$. All numerical moments entering the Lanczos construction are evaluated with the series truncation in Eq.~\eqref{2.3}. {We plot $1+K(t)$ for numerical convenience.} The vertical axis is plotted on a logarithmic scale.}
   	\label{fig:ktlist}
   \end{figure}

   Given the Lanczos coefficients, the Krylov complexity can be obtained by solving the discrete Schr\"odinger equation \eqref{2.12}. The time evolution of the Krylov complexity for various chemical potentials is shown in Fig.~\ref{fig:ktlist}.
   Note that the vertical axis is plotted on a logarithmic scale. We find that the complexity grows in time for all chemical potentials, but the curves do not collapse onto a single universal trajectory, in contrast to the non-positive-$\mu$ results of Ref.~\cite{He:2024hkw}. In particular, for relatively small $\mu$ (e.g. $\mu=1,10,20$), the complexity at a fixed time increases noticeably as $\mu$ increases. By contrast, for larger $\mu$ (e.g. $\mu=100,200$), the curves nearly coincide, suggesting that the complexity approaches an asymptotic behavior at large chemical potential. {An estimate of the crossover time scale $t^{\ast}$ from a continuum approximation is presented in Appendix~\ref{app:crossover}.}
   
   Although the curves in Fig.~\ref{fig:ktlist} may look close to exponential on a log scale over an intermediate time window, the true late-time behavior is quadratic. To explain this, we compare the large-$n$ asymptotics of our Lanczos coefficients with the $SL(2,\mathbb{R})$ results discussed in Ref.~\cite{Balasubramanian:2022tpr}. There, the authors studied the spread complexity for systems whose Hamiltonians are built from Lie-algebra generators.\footnote{In the state formalism, Krylov complexity is also referred to as spread complexity.} One representative case is a Hamiltonian constructed from $SL(2,\mathbb{R})$ generators:
   \begin{equation}
   	H=\alpha(L_{+}+L_{-})+\gamma L_{0}+\delta I,
   \end{equation}
   where $L_+$ and $L_-$ are the raising and lowering operators, $L_0$ is the Cartan generator, and $I$ is the identity operator. Here $\alpha$, $\gamma$ and $\delta$ are parameters. For $SL(2,\mathbb{R})$, the Lanczos coefficients are
   \begin{equation}\label{2.9}
   	a_n=\gamma(h+n)+\delta ,\qquad b_{n}=\alpha \sqrt{n(2h+n-1)}.
   \end{equation}
 where $h$ is another parameter.  This means that when $n$ is sufficiently large, both the Lanczos coefficients $a_n$ and $b_n$ exhibit linear growths. The corresponding complexity is given by:
   \begin{equation}\label{2.10}
   	K(t)=\frac{2h}{1-\frac{\gamma^{2}}{4\alpha^{2}}}\sinh^{2}\left (\alpha t\sqrt{1-\frac{\gamma^{2}}{4\alpha^{2}}}\right ).
   \end{equation}
   The mathematical structure of the spread complexity of quantum states is pretty similar to that of the Krylov complexity of operators. Therefore, the above formula can also be used as a reference to infer the late-time behavior of Krylov complexity from the asymptotic behaviors of the Lanczos coefficients. In particular, for $SL(2,\mathbb{R})$ the complexity is already given in Eq.~\eqref{2.10}.
   We now discuss the asymptotic behavior of $K(t)$ in four cases \cite{Balasubramanian:2022tpr}:
   \begin{enumerate}
   	\item If $\gamma^2 > 4\alpha^2$, $K(t)$ will exhibit periodic behavior, that is
   	\begin{equation}\label{3.122}
   		K(t)=\frac{2h}{\frac{\gamma^{2}}{4\alpha^{2}}-1}\sin^{2}\left (\alpha t\sqrt{\frac{\gamma^{2}}{4\alpha^{2}}-1}\right ).
   	\end{equation}
   	\item If $\gamma^2 < 4\alpha^2$, then at late times
   	\begin{equation}
   		K(t)\propto e^{2\abs{\alpha}\sqrt{1-\frac{\gamma^{2}}{4\alpha^{2}}}t}.
   	\end{equation}
   	\item If $\gamma^2 = 4\alpha^2$, then $K(t)$ shows quadratic growth,
   	\begin{equation}\label{case3}
   		K(t)=2h\alpha^{2}t^{2}.
   	\end{equation}
   	\item If $a_n = 0$, then $\gamma = 0$, and $K(t)$ grows exponentially,
   	\begin{equation}\label{3.155}
   		K(t)\propto e^{2\abs{\alpha}t},\qquad \text{for }t\rightarrow\infty. 
   	\end{equation}
   \end{enumerate}
It is observed that the Lanczos coefficients of the Schr\"odinger field theory we are considering take the following form in large $n$ (see equations \eqref{3.5} and \eqref{3.6}),
\begin{gather}
	a_{n}\sim -\frac{4}{\beta}n+c_{1},\\
	b_{n}\sim \frac{2}{\beta}n+c_{2},
\end{gather}
where $c_{1}$ and $c_{2}$ are constants. When $n$ is sufficiently large, we have:
\begin{equation}\label{2.18}
	\begin{aligned}
		b_{n}&\sim \frac{2}{\beta}\sqrt{\left (n+\frac{\beta}{2}c_{2}\right )^{2}}=\frac{2}{\beta}\sqrt{n^{2}+\left (\frac{\beta c_{2}}{2}\right )^{2}+n\beta c_{2}}\approx \frac{2}{\beta}\sqrt{n(n+\beta c_{2})}
	\end{aligned}
\end{equation}
Compared to \eqref{2.9}, this is equivalent to taking  $h = (\beta c_{2}+1)/2,\alpha=2/\beta$ and $\gamma=-4/\beta$, $\delta=c_{1}+\frac{2(\beta c_{2}+1)}{\beta}$ in \eqref{2.9}, and then we can write
\begin{equation}\label{2.19}
a_{n}=-\frac{4}{\beta}\left (\frac{\beta c_{2}+1}{2}+n\right )+c_{1}+\frac{2(\beta c_{2}+1)}{\beta}.
\end{equation}
Equations \eqref{2.18} and \eqref{2.19} have the same form as the $SL(2,\mathbb{R})$ Lanczos coefficients in \eqref{2.9}, so they are expected to share the same asymptotic behavior. From \eqref{2.18} and \eqref{2.19}, we have $\gamma^2 = 4\alpha^2$, and therefore the third case above implies quadratic growth of the Krylov complexity:
\begin{equation}\label{eq:K_quadratic}
	K(t)=2h\alpha^{2}t^{2}.
\end{equation}

{We now provide a possible physical interpretation of the difference between negative and positive chemical potentials for fermionic fields. Negative chemical potentials correspond to low fermion number density \cite{cook1995understanding}. In this regime, the Krylov complexity is relatively insensitive to $\mu$, which may explain why the curves for different negative chemical potentials nearly overlap \cite{He:2024hkw}. By contrast, for positive chemical potentials, the system has a high Fermi energy due to the larger fermion density, which can have a stronger effect on the Krylov complexity. As a result, increasing $\mu$ increases the complexity until it eventually approaches a limiting behavior. We leave a more detailed understanding of this effect for future work.}

Having established the numerical behavior in Sec.~\ref{sec3} for the specific Schr\"odinger field theory, we now show that these features follow more generally from the structure of the Wightman spectrum by turning to engineered spectra in Sec.~\ref{sec4}.

\section{Digression: Wightman power spectrum and Krylov complexity}\label{sec4}
{In this section (which constitutes the other main new component of this paper), we discuss the relationship between the Wightman power spectrum and Krylov complexity by analyzing engineered spectra with controlled decay and truncation.} The purpose of this section is to provide theoretical support for the numerical observations in Sec.~\ref{sec3} and to pinpoint which spectral features control which dynamical behaviors. More concretely: (i) in the first subsection of Sec.~\ref{sec4} (``The single-sided exponentially decaying Wightman power spectrum''), we show with an explicit solvable example that a single-sided exponentially decaying spectrum leads to the late-time \emph{quadratic} growth $K(t)\propto t^{2}$; this supports the quadratic late-time behavior inferred in Sec.~\ref{sec3} from the large-$n$ asymptotics of the Lanczos coefficients (cf. the discussion around Eqs.~\eqref{2.18}--\eqref{2.19}); (ii) in Sec.~\ref{sec4}, Subsec.~\ref{4.2} (``The Wightman power spectrum with exponential decay on both sides''), the $a=b$ analysis of the engineered two-sided spectrum \eqref{3.9} leads to Eq.~\eqref{3.15} and hence the asymptotic behavior $K(t)\propto\sinh^{2}\!\left(\pi t/\beta\right)$, supporting the statement that for sufficiently large positive $\mu$ the low-frequency part of \eqref{2.1} approaches the even spectrum \eqref{2.4}; (iii) the comparison plots around Fig.~\ref{3anbn} provide the mechanism-level explanation of why a \emph{single-sided} hard cutoff (as in $\Theta(\mu-\omega)$) produces a \emph{deflection} in $a_n,b_n$, whereas a \emph{symmetric two-sided} cutoff produces a \emph{plateau}.

A comment on methodology is also in order: the $SL(2,\mathbb{R})$ algebraic construction is employed in Sec.~\ref{sec3} to extract the late-time/large-$n$ asymptotics from the observed Lanczos coefficients, providing a compact and physically transparent route to the complexity growth. The present Sec.~\ref{sec4} then supplies an \emph{independent} spectral rationale by analyzing simplified/engineered Wightman power spectra with controlled decay and truncation properties. Finally, in Appendix~\ref{app:crossover}, we give a complementary (and more rigorous) derivation based on orthogonal polynomials, where the Wightman power spectrum plays the role of the weight function; this approach not only reproduces the same asymptotic scalings, but also naturally yields an estimate for the crossover location $n^{\ast}$ between the MP-dominated and Laguerre-dominated regimes. Therefore, the three viewpoints are consistent and should be viewed as complementary: an algebraic effective description (Sec.~\ref{sec3}), a spectral/engineered-spectrum characterization (Sec.~\ref{sec4}), and an orthogonal-polynomial derivation with crossover estimates (Appendix~\ref{app:crossover}).
\subsection{The single-sided exponentially decaying Wightman power spectrum}\label{subsec:single_sided_exp}
From the power spectrum \eqref{2.1} we see that the range of the frequency $\omega$ is $-\infty<\omega<\mu$, and it exhibits exponential decay in the high-frequency regime \footnote{The term ``high-frequency regime" here refers to the regime where $\abs{\omega}\gg1$.}. Since the frequency is bounded up by $\mu$, the Wightman power spectrum is a single-sided exponentially decaying function as $\omega\to-\infty$. We are going to explore the behavior of Krylov complexity corresponding to the single-sided exponentially decaying power spectrum. Therefore, we would like to construct some simple examples which are amenable to analytical calculations. To this end, we may consider the following power spectrum,
\begin{equation}\label{3.1}
	f^{W}(\omega)=2\pi \kappa e^{\kappa \omega},\qquad \omega\le 0,
\end{equation}
where $\kappa$ is a positive constant. The power spectrum is simple enough that we can calculate the moments analytically,
\begin{equation}
	\mu_{n}=\frac{1}{2\pi}\int_{-\infty}^{0}d\omega \omega^{n}f^{W}(\omega)=(-1)^{n}\kappa^{-n}\Gamma(n+1),
\end{equation}
where $\Gamma(\cdot)$ is the Gamma function. The Lanczos coefficients are
\begin{equation}\label{3.3}
	a_{n}=-\frac{1}{\kappa}(2n+1),\qquad b_{n}=\frac{n}{\kappa},
\end{equation}
where $n=1,2,3,\cdots$.  Correspondingly, the constants in the Lanczos coefficients \eqref{2.9} of $SL(2,\mathbb{R})$ take the following values,
\begin{equation}
	\alpha=\frac{1}{\kappa},\qquad h=\frac{1}{2},\qquad \gamma=-\frac{2}{\kappa}, \qquad  \delta=0.
\end{equation}
In this sense, the constants satisfy the above case 3 in \eqref{case3}. Thus, the Krylov complexity can be read off from Eq.~\eqref{2.10} as
\begin{equation}
	K(t)=\frac{t^{2}}{\kappa^{2}},
\end{equation}
which satisfies the quadratic growth in time. 

On the other hand, we can also solve the discrete Schr\"odinger equation \eqref{2.12} and then derive the Krylov complexity from its definition. Taking the Fourier transformation of the power spectrum \eqref{3.1} yields the autocorrelation function,
\begin{equation}
	\varphi_{0}(t)=\frac{\kappa}{\kappa+it}.
\end{equation}
Combining the Lanczos coefficients \eqref{3.3} in the discrete Schr\"odinger equation \eqref{2.12}, we obtain
\begin{equation}
	\varphi_{n}(t)=\frac{\kappa t^{n}}{(\kappa+it)^{n+1}}.
\end{equation}
Therefore, the Krylov complexity is given by
\begin{equation}
	K(t)=\sum_{n}n\abs{\varphi_{n}(t)}^{2}=\sum_{n}n\kappa^{2}t^{2n}(\kappa^{2}+t^{2})^{-1-n}=\frac{t^{2}}{\kappa^{2}}.
\end{equation}
As expected, the two methods yield the same results that the Krylov complexity exhibits quadratic growth in time.

\subsection{The Wightman power spectrum with exponential decay on both sides}\label{4.2}

We can also consider a more general power spectrum that is close to \eqref{2.1},
\begin{equation}\label{3.9}
	f^{W}(\omega)=\mathcal{N}\frac{e^{a\omega}}{1+e^{(a+b)\omega}},\qquad  b\ge a>0,
\end{equation}
where $\mathcal{N}$ is a normalization constant. \footnote{For $a\ge b>0$, it can be dealt with similarly. Please refer to the Appendix \ref{appa}.} This power spectrum has some properties:
\begin{itemize}
	\item As $\omega \rightarrow +\infty$, $f^W(\omega)$ asymptotically approaches $e^{-b\omega}$. Conversely, as $\omega \rightarrow -\infty$, $f^W(\omega)$ tends toward $e^{a\omega}$.
	\item If $a = b$, then \eqref{3.9} is an even function and exhibits exponential decay as $\omega\to\pm\infty$. This spectrum is consistent with the universal operator-growth conjecture \cite{Parker:2018yvk}, and its Krylov complexity therefore shows exponential growth. In particular, for $a=b=\beta/2$, \eqref{3.9} reduces to \eqref{2.4}. This suggests that as the chemical potential $\mu$ increases, the Krylov complexity associated with \eqref{2.1} gradually approaches the exponential-growth behavior set by \eqref{2.4}.
	\item If $b \rightarrow +\infty$, then \eqref{3.9} reduces to \eqref{3.1}, and thus its Krylov complexity will exhibit quadratic growth.
\end{itemize}
\subsubsection{$a=b$}
For $a = b$, the normalization constant $\mathcal{N}$ is
\begin{equation}
	\mathcal{N}=4a.
\end{equation}
Therefore, the analytical results of the moments are
\begin{equation}
	\mu_{n}=\frac{2^{-2n-1}((-1)^{n}+1)a^{-n}(\zeta(n+1,1/4)-\zeta(n+1,3/4))\Gamma(n+1)}{\pi},
\end{equation}
 where $\zeta(\cdot,\cdot)$ is the generalized Riemann zeta function. The factor $(-1)^n + 1$ in the numerator indicates that for odd $n$, $\mu_n = 0$. In this case, only the Lanczos coefficients $b_n$ are nonzero,
\begin{equation}\label{3.12}
	b_{n}=\frac{n\pi}{2a}.
\end{equation}
Taking $a = \beta / 2$, the power spectrum \eqref{3.9} reduces to \eqref{2.4}, and the Lanczos coefficients $b_n$ satisfy $\beta b_n =n\pi$. In the previous Figure \ref{fig:bnlist} the red dashed line represents $\beta b_n =n\pi$, indicating that as $\mu$ increases, the power spectrum $\eqref{2.1}$ indeed gradually approaches \eqref{2.4}. A natural idea is that as the chemical potential increases, the Krylov complexity given by \eqref{2.1} will approach the Krylov complexity given by \eqref{2.4} as well. Solving the discrete Schr\"odinger equation with $b_n$ in \eqref{3.12} yields
\begin{equation}
	\varphi_{n}(t)=\sech\left(\frac{\pi t}{2a}\right)\tanh^n\left(\frac{\pi t}{2a}\right).
\end{equation}
Therefore, the Krylov complexity becomes
\begin{equation}
	K(t)=\sum_{n}n\abs{\varphi_{n}(t)}^{2}=\sinh^2\left(\frac{\pi t}{2a}\right).
\end{equation}
Taking $a=\beta/2$, we have 
\begin{equation}\label{3.15}
	K(t)=\sinh^{2}\left(\frac{\pi t}{\beta}\right).
\end{equation}
\begin{figure}
	\centering
	\includegraphics[width=0.6\linewidth]{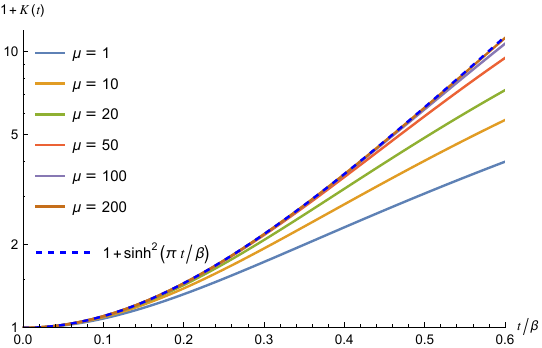}
	\caption{Comparison of Krylov complexities with different chemical potentials $\mu = 1, 10, 20, 50,$ $100$ and $200$. The blue dashed line represents the asymptotic behavior $1+K(t) = 1 + \sinh^2 (\pi t / \beta)$. The curve for the complexity with $\mu = 200$ approaches the asymptotic behavior closely. }
	\label{fig:ktt}
\end{figure}

In Figure \ref{fig:ktt}, we compare the Krylov complexity in \eqref{3.15} with those for different chemical potentials as discussed in Figure \ref{fig:ktlist}. The blue dashed line in this figure represents the asymptotic behavior \eqref{3.15}. We find that as the chemical potential increases, the Krylov complexities approach the result in \eqref{3.15}. When the chemical potential is $\mu = 200$, the corresponding curve is already very close to \eqref{3.15}. This figure supports the idea that, as $\mu$ increases, the Krylov complexity for \eqref{2.1} approaches that for the even spectrum \eqref{2.4}.

\begin{figure}
	\centering
	\subfigure[Lanczos coefficients $a_{n}$ for different Wightman power spectra.]{\includegraphics[width=0.9\textwidth]{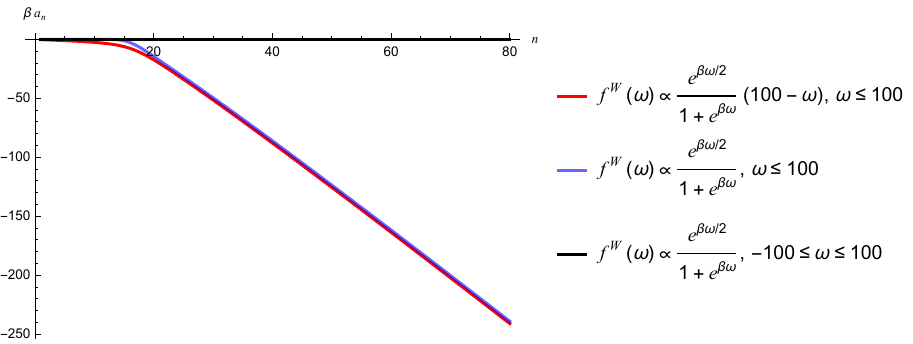}\label{3an}}
	\vfill
	\subfigure[Lanczos coefficients $b_{n}$ for different Wightman power spectra.]{\includegraphics[width=0.9\textwidth]{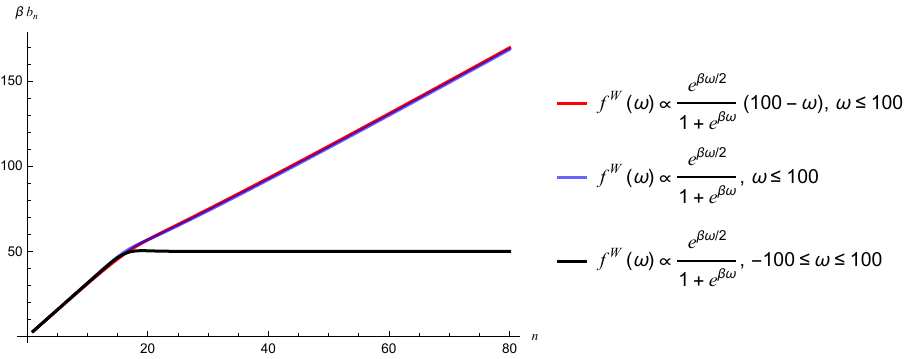}\label{3bn}}
	\caption{{Lanczos coefficients $a_{n}$ and $b_{n}$ for different Wightman power spectra $f^{W}(\omega)$, highlighting the role of frequency cutoffs/truncations. The red and blue curves correspond to a \emph{single-sided} truncation (fit/truncation window $\omega\le 100$, i.e. $\omega\in(-\infty,100]$), while the black curves correspond to a \emph{symmetric two-sided} truncation (window $|\omega|\le 100$, i.e. $\omega\in[-100,100]$). All numerical moments are computed with the same series truncation $\sum_{k=0}^{200}(\cdots)$.}}
	\label{3anbn}
\end{figure}

Now we discuss how the Lanczos coefficients depend on frequency cutoffs in the Wightman power spectrum. In Figs.~\ref{fig:anlist} and \ref{fig:bnlist}, the spectrum \eqref{2.1} has a single-sided support $\omega\le \mu$, and we observe clear deflections in both $a_n$ and $b_n$. By contrast, when the spectrum is truncated symmetrically, i.e. $|\omega|\le \Lambda$, the Lanczos coefficient $b_n$ often develops a plateau at large $n$ \cite{Camargo:2022rnt,He:2024hkw} (see, for example, the black curve in Fig.~\ref{3bn}).

This motivates the following question: how are the observed deflections/plateaus in the Lanczos coefficients related to the cutoff structure of the spectrum?\footnote{Here ``deflection'' refers to a visible bending in the curve (e.g. the red and blue curves in Fig.~\ref{3bn}), while ``plateau'' refers to the emergence of an approximately horizontal segment at large $n$ (e.g. the black curve in Fig.~\ref{3bn}).} A natural hypothesis is that the Heaviside factor $\Theta(\mu-\omega)$ in \eqref{2.1} acts as a single-sided hard cutoff and produces a deflection, whereas a symmetric two-sided cutoff $|\omega|\le \Lambda$ produces a plateau. We verify these expectations numerically in Fig.~\ref{3anbn}.

In Fig.~\ref{3anbn}, we compare the Lanczos coefficients $a_n$ and $b_n$ obtained from three different power spectra. The first spectrum is $f^W(\omega) \propto \frac{e^{\beta \omega/2}}{1 + e^{\beta \omega}} (100 - \omega)$ with $\omega \leq 100$, which corresponds to setting $\mu = 100$ in Eq.~\eqref{2.1}; the corresponding Lanczos coefficients are shown by the red curves. The second spectrum is $f^W(\omega) \propto \frac{e^{\beta \omega/2}}{1 + e^{\beta \omega}}$ with $\omega \leq 100$, i.e. Eq.~\eqref{2.4} truncated at $\omega = 100$; the corresponding Lanczos coefficients are shown by the blue curves. The third spectrum is $f^W(\omega) \propto \frac{e^{\beta \omega/2}}{1 + e^{\beta \omega}}$ with $-100 \leq \omega \leq 100$, i.e. Eq.~\eqref{2.4} truncated at both $\omega = \pm 100$; the corresponding Lanczos coefficients are shown by the black curves.

We observe that for single-sided truncations ($\omega\leq 100$), the Lanczos coefficients obtained from \eqref{2.1} and \eqref{2.4} both exhibit clear deflections, and the deflection locations are close to each other (around $n\approx16$; see the red and blue curves). By contrast, for the symmetric truncation ($|\omega|\leq 100$), the Lanczos coefficient $b_n$ develops a plateau, and the onset of the plateau is close to the deflection locations of the single-sided cases.

\subsubsection{\texorpdfstring{$a\neq b$}{a!=b}}
\begin{figure}[b]
	\centering
	\subfigure[Linear decrease of $\beta a_{n}$ against $n$ for various $b$. ]{\includegraphics[width=0.48\textwidth]{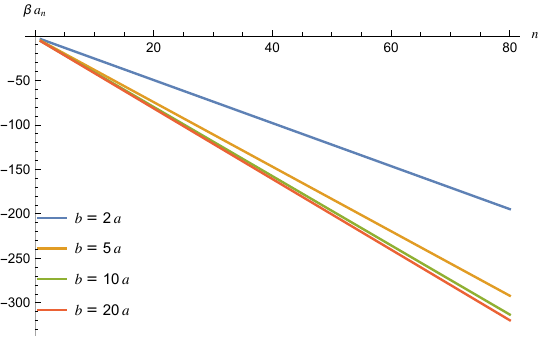}}~~
	\subfigure[Linear increase of $\beta b_{n}$ against $n$ for various $b$. ]{\includegraphics[width=0.48\textwidth]{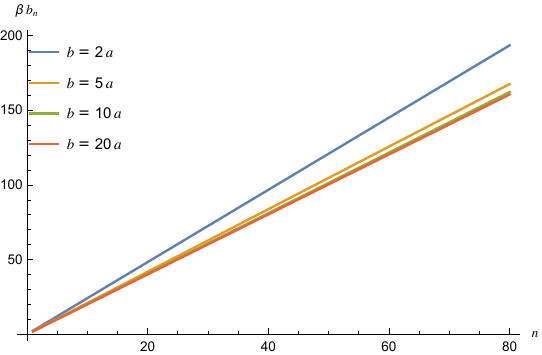}}
	\caption{{Lanczos coefficients $a_n$ and $b_n$ versus $n$ for various $b$ (with fixed $a=\beta/2$). All numerical moments entering the Lanczos construction are evaluated with the series truncation $\sum_{k=0}^{200}(\cdots)$ like Eq.~\eqref{2.3}.}}
	\label{rhoab}
\end{figure}

If $a\neq b$, the power spectrum \eqref{3.9} has different decay rates as $\omega \rightarrow \pm\infty$. We now study the resulting behaviors of the Lanczos coefficients $a_n$ and $b_n$. For definiteness, we consider $b>a$; the normalization constant in \eqref{3.9} becomes
\begin{equation}
	\mathcal{N}=2(a+b)\sin\left (\frac{b\pi}{a+b}\right ).
\end{equation}
We can get the moments $\mu_{n}$ as
\begin{equation}
	\begin{aligned}
		\mu_{n}=&2^{-n} (a + b)^{-n} \Gamma(1 + n) \sin\left(\frac{b \pi}{a + b}\right) \times \\
		&\left( \zeta\left(1 + n, \frac{b}{2 (a + b)}\right) + (-1)^n \left( \zeta\left(1 + n, \frac{a}{2 (a + b)}\right) - \zeta\left(1 + n, \frac{2a + b}{2 (a + b)}\right) \right) \right. \\
		&\left. - \zeta\left(1 + n, 1 - \frac{a}{2 (a + b)}\right) \right).
	\end{aligned}
\end{equation}
This formula is too complicated to calculate analytically, therefore, we choose to compute it numerically and without loss of generality we set $a = \beta / 2$. Using the moment method, we obtain the Lanczos coefficients for different values of $b$, see Figure. \ref{rhoab}.
The results in the figure indicate that when $b > a$, the Lanczos coefficients $a_n$ and $b_n$ both exhibit linear behavior. Specifically, for a fixed $a$, the larger the value of $b$, the smaller the slopes of the Lanczos coefficients are. The fitted linear behaviors of these Lanczos coefficients are,
\begin{gather}
	b=2a:\qquad \beta a_{n}=-1.21 - 2.42 n,\qquad \beta b_{n}=2.42 n,\\
	b=5a:\qquad \beta a_{n}=-1.81 - 3.63 n,\qquad \beta b_{n}=2.09n,\\
	b=10a:\qquad \beta a_{n}=-1.95 - 3.89 n,\qquad \beta b_{n}=2.03n,\\
	b=20a:\qquad \beta a_{n}=-1.99 - 3.97 n,\qquad \beta b_{n}=2.01n.
\end{gather}

{This indicates that as $b$ increases, the Lanczos coefficients approach the asymptotic scalings $\beta a_n\to-4n$ and $\beta b_n\to 2n$. According to Eqs.~\eqref{3.122}--\eqref{3.155}, this implies that the Krylov complexity exhibits quadratic growth in the large-$b$ limit.}

\section{Conclusions}\label{sec5}
{In this work we investigate the evolution of Krylov complexity in Schr\"odinger field theory at \emph{positive} chemical potential $\mu$, and support our numerical observations using engineered Wightman power spectra with controlled frequency truncations. In the numerical part, all moments entering the Lanczos construction are evaluated with a finite series truncation $\sum_{k=0}^{200}(\cdots)$; this truncation is a numerical regulator (not a physical cutoff) and is used only to extract the Lanczos data in a controllable way, while the qualitative features we emphasize---the single-sided truncation of $f^{W}(\omega)$ at $\omega=\mu$ and the associated crossover/deflection in $a_n,b_n$---are explained by independent analytic arguments.}

Our main result is a dynamical transition induced by the single-sided spectral cutoff at $\omega=\mu$: as $\mu$ becomes positive and increases, the Lanczos coefficients develop a characteristic deflection/two-stage structure. In particular, $b_{n}$ exhibits an early-stage linear growth with slope $\pi/\beta$ and then crosses over to an asymptotic slope $2/\beta$, while $a_{n}$ changes from near-zero values to an asymptotic linear descent with slope $-4/\beta$. This behavior is qualitatively different from the non-positive-$\mu$ case \cite{He:2024hkw} and provides a spectral interpretation of how chemical potential reshapes operator growth.

We also emphasize that the asymptotic behaviors inferred in Sec.~\ref{sec3} are supported from two complementary viewpoints: (i) an algebraic effective description based on the $SL(2,\mathbb{R})$ construction (used in Sec.~\ref{sec3} to connect large-$n$ Lanczos asymptotics to complexity growth), and (ii) a spectral/orthogonal-polynomial characterization (Sec.~\ref{sec4} and Appendix~\ref{app:crossover}) that relates deflections/plateaus of $a_n,b_n$ to hard cutoffs/truncation windows in $f^{W}(\omega)$ and yields an estimate for the crossover location $n^{\ast}$. These viewpoints are consistent and together provide a coherent picture of the late-time dynamics.

Using engineered spectra, we further illustrate how the \emph{truncation window} affects complexity growth: single-sided exponential decay (and a single-sided hard cutoff) leads to quadratic growth at late times, while a two-sided exponentially decaying spectrum (approximately recovered at large $\mu$) restores an exponential-growth regime consistent with maximal-chaos-like expectations. Moreover, the comparison in Fig.~\ref{3anbn} shows that a single-sided truncation produces a deflection in the Lanczos coefficients, whereas a symmetric two-sided truncation produces a plateau, and an asymmetric two-sided truncation can generate two inflection points.

To summarize the underlying mechanism in one sentence: the hard one-sided cutoff at $\omega=\mu$ makes the effective large-frequency tail of $f^{W}(\omega)$ intrinsically \emph{single-sided}, which drives the Lanczos data away from the MP-type two-sided exponential regime and into the Laguerre-type asymptotics ($b_n\sim 2n/\beta$, $a_n\sim -4n/\beta$), and this change of asymptotic Lanczos growth enforces a late-time quadratic complexity $K(t)\propto t^{2}$.

{Finally, we stress that the operators considered in this work are already \emph{non-Hermitian} in general, and consequently the Wightman power spectrum is not required to be an even function of $\omega$. Our results show that the combination of spectral \emph{asymmetry} (non-Hermiticity) and hard cutoffs/truncations can leave sharp signatures in Lanczos data (deflections/plateaus) and in the resulting complexity growth. It would be interesting to further explore these signatures in more general non-Hermitian settings (e.g. different operator choices) and in interacting Schr\"odinger-invariant theories, which we leave for future work.}

\acknowledgments
This work was partially supported by the National Natural Science Foundation of China (Grants No.~12175008 and 12165009), the Hunan Natural Science Foundation (Grants No.~2023JJ30487 and 2022JJ40340),
the Key Project of Sichuan Science and Technology Education Joint Fund (No.~25LHJJ0097), and the Sichuan Natural Science Foundation (Grant No.~2026NSFSC0746).

\appendix

{\section{Orthogonal-polynomial argument and crossover estimates}\label{app:crossover}

In this appendix, we provide a theoretical derivation that (i) predicts the location of the dynamical transition point of the Lanczos coefficients, and (ii) explains why the asymptotic growth of the Krylov complexity must be quadratic.

To rigorously characterize the complexity growth and spectral properties, we employ the mapping between the Krylov operator dynamics and the theory of orthogonal polynomials. Let the basis operators in the Krylov space be generated by polynomials $P_n(\mathcal{L})$ acting on the initial normalized operator $\left|O_0\right)$ \cite{Muck:2022xfc}:
\begin{equation}
	\left|O_n\right) = P_{n}(\mathcal{L})\left|O_{0}\right) \,.
\end{equation}
The Lanczos three-term recurrence relation for the orthonormal basis operators,
\begin{equation}
	\mathcal{L}\left|O_{n}\right) = b_{n}\left|O_{n-1}\right) + a_{n}\left|O_{n}\right) + b_{n+1}\left|O_{n+1}\right) \,,
\end{equation}
implies that the polynomials $P_n(\omega)$ satisfy the same algebraic recurrence in the frequency domain. Replacing the superoperator $\mathcal{L}$ with the scalar spectral variable $\omega$:
\begin{equation}\label{app:OP_rec}
	\omega P_{n}(\omega) = b_{n}P_{n-1}(\omega) + a_{n}P_{n}(\omega) + b_{n+1}P_{n+1}(\omega) \,.
\end{equation}
The orthonormality of the Krylov basis, $\delta_{nm} = (O_n|O_m)$, enforces the orthogonality of these polynomials with respect to the Wightman power spectrum $f_W(\omega)$ as the weight function:
\begin{equation}
	\frac{1}{2\pi}\int_{-\infty}^{\infty} f^{W}(\omega)P_{n}(\omega)P_{m}(\omega)d\omega = \delta_{nm} \,.
\end{equation}
The behavior of $a_n$ and $b_n$ is entirely determined by the weight function $f^W(\omega)$. 

In this work, we derived the Wightman power spectrum for the fermionic field with a chemical potential $\mu$ which is given by:
\begin{equation} \label{app:full_spectrum}
	f^W(\omega) \sim (\mu-\omega)\dfrac{e^{\beta\omega/2}}{1+e^{\beta\omega}} \Theta(\mu - \omega) \,.
\end{equation}
The Heaviside function $\Theta(\mu-\omega)$ imposes a hard cutoff at the chemical potential. This spectrum naturally interpolates between two distinct regimes that govern the operator growth at different stages. At early times (small $n$), the dynamics is dominated by the low-frequency part of the spectrum ($|\omega| \ll \mu$). In this regime, the linear term is slowly varying $(\mu-\omega \approx \mu)$, and the spectrum is dominated by the thermal factor:
\begin{equation}\label{app:thermal_weight}
	f_W(\omega) \sim \sech\left(\frac{\beta \omega}{2}\right) \,.
\end{equation}
This weight function corresponds to the unique measure of the Meixner--Pollaczek (MP) polynomials $P_n^{(\lambda)}(x;\phi)$. The MP polynomials are characterized by the weight function $w(x;\lambda,\phi)$ and the three-term recurrence relation as follows \cite{chihara2011introduction}:
\begin{gather}
	w(x;\lambda,\phi)=e^{(2\phi-\pi)x}\abs{\Gamma(\lambda+ix)^{2}},\\
	(n+1)P_{n+1}^{(\lambda)}(x;\phi)-2[x\sin\phi+(n+\lambda)\cos\phi]P_{n}^{(\lambda)}(x;\phi)+(n+2\lambda-1)P^{(\lambda)}_{n-1}(x;\phi)=0.
\end{gather}
Additionally, the MP polynomials satisfy
\begin{equation}
	\int_{-\infty}^{\infty}P_{n}^{(\lambda)}(x;\phi)P^{(\lambda)}_{m}(x;\phi)w(x;\lambda,\phi)dx=h_{n}\delta_{nm},\qquad h_{n}=\dfrac{2\pi\Gamma(n+2\lambda)}{n!(2\sin\phi)^{2\lambda}}.
\end{equation}
Specifically, by setting $\phi =\frac{\pi }{2}$ and $\lambda =\frac{1}{2}$, the weight function $w(x;\lambda,\phi)$ of the MP polynomials reduces to:
\begin{equation} \label{app:mp_weight}
	w(x) = \frac{\pi}{\cosh(\pi x)} ,
\end{equation}
and the three-term recurrence relation becomes
\begin{equation}
	(n+1)P_{n+1}(x)-2xP_{n}(x)+nP_{n-1}(x)=0.
\end{equation}
Equation \eqref{app:mp_weight} has the same form as \eqref{app:thermal_weight}; therefore, the Lanczos coefficients can be obtained from the three-term recurrence relation of the MP polynomials. To read out these coefficients, we need to normalize the MP polynomials first. Let $\mathcal{P}_n(x)=\frac{1}{\sqrt{h_n}}P_{n}(x)$ denote the normalized orthogonal polynomials, then
\begin{equation}
	x\mathcal{P}_{n}(x)=\dfrac{n+1}{2}\mathcal{P}_{n+1}(x)+\dfrac{n}{2}\mathcal{P}_{n-1}(x).
\end{equation}
To exactly match the power spectrum, we set $\pi x=\dfrac{\beta\omega}{2}$, thus:
\begin{equation}\label{app:mp_rec_w}
	\omega\mathcal{P}_{n}(x(\omega))=\dfrac{(n+1)\pi}{\beta}\mathcal{P}_{n+1}(x(\omega))+\dfrac{n\pi}{\beta}\mathcal{P}_{n-1}(x(\omega)).
\end{equation}
Comparing \eqref{app:mp_rec_w} and \eqref{app:OP_rec}, we obtain
\begin{equation}
	a_{n}=0,\qquad b_{n}=\dfrac{n\pi}{\beta}.
\end{equation}

In the late-time limit, the Wightman power spectrum is dominated by the high-frequency region. Hence, it is approximately given by:
\begin{equation}
	f^{W}(\omega)\propto (\mu-\omega)e^{\beta\omega/2} \qquad \omega\in(\infty,\mu].
\end{equation}
Such a power spectrum can be related to generalized Laguerre polynomials $L^{(\alpha)}_{n}(x)$, with the weight function given by
\begin{equation}
	W(x)=\dfrac{2\pi}{\Gamma(\alpha+1)}x^{\alpha}e^{-x},\qquad  {\blue x\in} [0,\infty),
\end{equation}
and the three-term recurrence relation is
\begin{equation}
	\begin{aligned}
		xL^{(\alpha)}_{n}(x)=&-\sqrt{(n+1)(n+1+\alpha)}L^{(\alpha)}_{n+1}(x)\\
		&+(2n+\alpha+1)L^{(\alpha)}_{n}(x)-\sqrt{n(n+\alpha)}L^{(\alpha)}_{n-1}(x).
	\end{aligned}
\end{equation}
To match the power spectrum with the weight function precisely, let $x= c(\mu-\omega)$, where $c$ is a constant. Then we have 
\begin{equation}
	f^{W}(\omega(x))\sim e^{\frac{\beta\mu}{2}}\left(\dfrac{x}{c}\right)e^{-\frac{\beta x}{2c}}\Rightarrow c=\dfrac{\beta}{2},\alpha=1.
\end{equation}
The three-term recurrence relation becomes
\begin{equation}
	\begin{aligned}
		\omega L^{(1)}_{n}(x(\omega))=&\left (\mu-\dfrac{4n+4}{\beta}\right )L^{(1)}_{n}(x(\omega))+\dfrac{2\sqrt{(n+1)(n+2)}}{\beta}L^{(1)}_{n+1}(x(\omega))\\
		&+\dfrac{2\sqrt{n(n+1)}}{\beta}L_{n-1}^{(1)}(x(\omega)).
	\end{aligned}
\end{equation}
For large $n$, the Lanczos coefficients read off from this equation are:
\begin{equation}
	a_{n}=-\dfrac{4n}{\beta}-\dfrac{4}{\beta}+\mu,\qquad b_{n}=\dfrac{2}{\beta}\sqrt{n(n+1)}\approx\dfrac{2n}{\beta}+\dfrac{1}{\beta}.
\end{equation}

{It can be readily seen that the slopes of $a_n$ and $b_n$ agree with the numerical results in the maintext, though the intercepts need not. The latter discrepancy originates from the fact that at low frequencies the exact Wightman power spectrum deviates from the Laguerre-polynomial weight function used in this asymptotic matching. Nevertheless, the combination $a_{n}+2b_{n}=\mu-\dfrac{2}{\beta}$ remains valid within the present Laguerre asymptotics (and is also in agreement with our numerical Lanczos data). This distinction is mathematically well-founded and is consistent with the asymptotic theory of orthogonal polynomials \cite{szeg1939orthogonal,magnus1987asymptotic}.}

We now proceed to determine the turning point of the Lanczos coefficients. In quantum mechanics, the energy eigenstates of the harmonic oscillator satisfy the following recurrence relation:
\begin{equation}
	x\psi_{n}(x)=\sqrt{\dfrac{\hbar}{2\mu\omega}}[\sqrt{n+1}\psi_{n+1}(x)+\sqrt{n}\psi_{n-1}(x)].
\end{equation}
This indicates that in the energy representation, the operator $x$  admits a tridiagonal matrix representation. Likewise, an analogous interpretation applies to the recurrence relation of our orthogonal polynomials, namely that the operator $\omega$  can be represented as a tridiagonal matrix:
\begin{equation}
	\begin{pmatrix}
		a_0 & b_1 & 0 & 0 & \cdots & 0 \\
		b_1 & a_1 & b_2 & 0 & \cdots & 0 \\
		0 & b_2 & a_2 & b_3 & \cdots & 0 \\
		0 & 0 & b_3 & a_3 & \ddots & \vdots \\
		\vdots & \vdots & \vdots & \ddots & \ddots & \cdots \\
		0 & 0 & 0 & \cdots & \cdots & \cdots
	\end{pmatrix}.
\end{equation}
Therefore, the part dominated by MP polynomials can be regarded as lying in a subspace. Assuming that MP polynomials are no longer applicable when $n>n^{\ast}$, a turning point will emerge. To determine $n^{\ast}$, we employ the Gershgorin circle theorem \cite{golub2013matrix}:

\textit{Suppose $A\in \mathbb{R}^{n\times n}$ is symmetric and that $Q\in \mathbb{R}^{n\times n}$ is orthogonal. If $Q^{T}AQ=D+F$ where $D=\text{diag}(d_{1},\cdots,d_{n})$ and $F$ has zero diagonal entries, then
	\begin{equation}
		\lambda(A)\subseteq \bigcup_{i=1}^{n}[d_{i}-r_{i},d_{i}+r_{i}]
	\end{equation}
	where $\lambda(A)$ denotes the set of eigenvalues of $A$ and $r_{i}=\sum\limits_{j=1}^{n}\abs{f_{ij}}$ for $i=1,2,\cdots,n$.}

According to this theorem, the MP-dominated part corresponds to:
\begin{equation}
	\abs{\omega}\le b_{n^{\ast}-1}+b_{n^{\ast}}\approx \mu,
\end{equation}
then we obtain
\begin{equation}
	n^{\ast}\approx\dfrac{\beta}{2\pi}{\left(\mu+\dfrac{\pi}{\beta}\right)}\sim \dfrac{\beta\mu}{2\pi}.
\end{equation}
We have computed  $\dfrac{\beta\mu}{2\pi}$
for $\beta=1$ and $\mu=1,10,20,50,100,200 $, and the results are shown below:
\begin{gather}
	\mu=1,\qquad \dfrac{\beta\mu}{2\pi}= 0.159;\\
	\mu=10,\qquad \dfrac{\beta\mu}{2\pi}= 1.59;\\
	\mu=20,\qquad \dfrac{\beta\mu}{2\pi}= 3.18;\\
	\mu=50,\qquad \dfrac{\beta\mu}{2\pi}= 7.96;\\
	\mu=100,\qquad \dfrac{\beta\mu}{2\pi}= 15.9;\\
	\mu=200,\qquad \dfrac{\beta\mu}{2\pi}= 31.8.
\end{gather}
Compared with Figure 1 in this paper, we can see that the positions of the turning points in the figure agree well with our calculated results here. 

In Section 4.2.1 of the maintext, we explained that the Krylov complexity initially evolves as $\sinh^{2}{\left(\dfrac{\pi t}{\beta}\right)}$. Now that we know the region dominated by MP polynomials, we can determine roughly when the Krylov complexity deviates from $\sinh^{2}{\left(\dfrac{\pi t}{\beta}\right)}$. 

To investigate the dynamics of the Krylov complexity, we employ the continuum limit approximation. We introduce the continuous coordinate $x_{n}=\varepsilon n$ and map the discrete quantities to continuous functions:
\begin{equation}
	a_{n}=a(x_{n})\approx 0,\qquad b_{n}=b(x_{n}),\qquad \varphi_{n}(t)=\varphi(x_{n},t)\,.
\end{equation}
Here the approximation $a(x_n)\approx 0$ reflects the fact that we focus on the MP-polynomial-dominated regime, in which the corresponding three-term recurrence gives $a_n=0$ (see the discussion above). Away from this regime (after the crossover), $a_n$ can deviate from zero, and its effect is subleading in the present early-time/continuum analysis.
With this identification, we can expand the discrete recurrence relation. Applying a Taylor expansion to the nearest-neighbor terms yields the time evolution equation in the first order of the lattice spacing $\varepsilon$:
\begin{equation}
	\begin{aligned}
		\partial_{t}\varphi(x_{n},t)&=b(x_{n})\left [\varphi(x_{n},t)-\dfrac{\partial \varphi(x_{n},t)}{\partial x_{n}}\varepsilon\right ]
		-\left [b(x_{n})+\dfrac{\partial b(x_{n})}{\partial x_{n}}\varepsilon\right ]\left [\varphi(x_{n},t)+\dfrac{\partial \varphi(x_{n},t)}{\partial x_{n}}\varepsilon\right ]\\
		&=-2b(x_{n})\dfrac{\partial\varphi(x_{n},t)}{\partial x_{n}}\varepsilon-\dfrac{\partial b(x_{n})}{\partial x_{n}}\varphi(x_{n},t)\varepsilon,
	\end{aligned}
\end{equation}
which simplifies to the standard advection-diffusion-like partial differential equation:
\begin{equation}
	\partial_{t}\varphi(x,t)=-2b(x)\partial_{x}\varphi(x,t)\varepsilon-\partial_{x}b(x)\varphi(x,t)\varepsilon.
\end{equation}

This partial differential equation can be solved using the method of characteristics. We consider the total derivative of the wave function with respect to time along a trajectory $x(t)$, which is given by the chain rule:
\begin{equation}
	\dfrac{d}{dt}\varphi(x,t)=\partial_{x}\varphi(x,t)\dfrac{dx}{dt}+\partial_{t}\varphi(x,t)
\end{equation}
By comparing the total derivative with the derived PDE, we identify the equation governing the characteristic curves, which represents the velocity of the information propagation in the Krylov space:
\begin{equation}
	\dfrac{dx}{dt}=2b(x)\varepsilon
\end{equation}
In the early growth regime, the Lanczos coefficients exhibit linear behavior $b(x)=\dfrac{\pi x}{\beta}$. Substituting this into the characteristic equation and integrating yields an exponential growth for the position $x(t)$. The scrambling time $t^{\ast}$ is then defined as the time required for the wave packet to propagate to the effective edge of the Krylov chain determined by the chemical potential $\mu$:
\begin{equation}
	x(t)\sim e^{2\pi t/\beta}\Rightarrow t^{\ast}\sim \dfrac{\beta}{2\pi}\ln\dfrac{\beta\mu}{2\pi}
\end{equation}
Based on this logarithmic relation, the specific scrambling times for various values of the chemical potential $\mu$ are estimated as follows:
\begin{gather}
	\mu=1,\qquad t^{\ast}\approx -0.293,\\
	\mu=10,\qquad t^{\ast}\approx 0.074,\\
	\mu=20,\qquad t^{\ast}\approx 0.184,\\
	\mu=50,\qquad t^{\ast}\approx 0.330,\\
	\mu=100,\qquad t^{\ast}\approx 0.440,\\
	\mu=200,\qquad t^{\ast}\approx 0.551.
\end{gather}
We note that this timescale is also consistent with the numerical observation in Fig.~3 of the paper. In particular, for $\mu=1$ we obtain $t^{\ast}<0$, which implies that from the very beginning the Krylov complexity does not follow the early-time growth $\sinh^2\left(\frac{\pi t}{\beta}\right)$.

As schematically illustrated in Fig.~\ref{fig:K_asymp_Lanczos}, we denote $x^{\ast}=\varepsilon n^{\ast}$. When $x<x^{\ast}$, the MP-polynomial regime dominates, while for $x>x^{\ast}$ the dynamics is controlled by the Laguerre-polynomial regime. The red line in the figure represents a (purely schematic) characteristic line, not the actual one. Along this line, the crossing $x=x^{\ast}$ occurs at $t=t^{\ast}$. The two reference values $x=x^{\ast}$ and $t=t^{\ast}$ divide the $t$--$x$ plane into four regions labeled 1,2,3,4. At the beginning of the evolution, the wave function is mainly supported in region 1, so the Krylov complexity is almost entirely controlled by the MP-polynomial regime and follows the complexity evolution associated with a two-sided exponentially decaying spectrum~\cite{He:2024xjp}. As time increases, the wave packet gradually propagates into the domain $x>x^{\ast}$ (Laguerre-dominated), and therefore the late-time asymptotics approaches the behavior governed by the Laguerre-polynomial regime.

\begin{figure}[ht]
	\centering
	\includegraphics[width=0.5\linewidth]{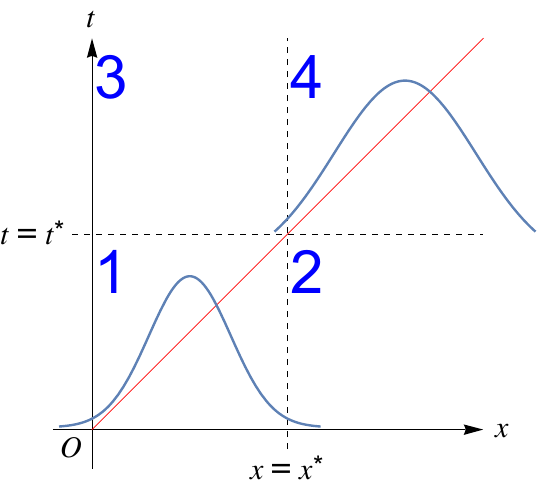}
	\caption{A schematic illustration that the asymptotic behavior of the Krylov complexity is controlled by the asymptotic behavior of the Lanczos coefficients.}
	\label{fig:K_asymp_Lanczos}
\end{figure}

} 

\section{The Lanczos coefficients of \texorpdfstring{$f^W(-\omega)$}{fW(-omega)}}\label{appa}

In Section \ref{4.2}, for the power spectrum \eqref{3.9}, we only considered the case where $b\ge a>0$. In principle, we can also consider the case where $a\geq b>0$. This is equivalent to considering a power spectrum that is symmetric to \eqref{3.9} about $\omega=0$, namely $\tilde f^{ W}(\omega)=f^W(-\omega)$. Generally speaking, the moments of $\tilde f^{ W}(\omega)$ are related to the moments of $f^W(\omega)$ in the following way
\begin{equation}
	\tilde\mu_n=(-1)^n\mu_n,
\end{equation}
where $\tilde\mu_n$ is derived from $\tilde f^{ W}(\omega)$. For $f^W(\omega)$, we can write down its Lanczos coefficients as
\begin{eqnarray}
	a_{1}&=&-\frac{\mu_{1}^{3}-2\mu_{1}\mu_{2}+\mu_{3}}{\mu_{1}^{2}-\mu_{2}},\label{a2}\\
		a_{2}&=&\frac{
			\mu_3^3 - 2\mu_2\mu_3\mu_4 
			- \mu_1\mu_2 (\mu_2^3 + 2\mu_3^2 - 2\mu_2\mu_4) 
			- \mu_1^3 (\mu_3^2 + 2\mu_2\mu_4) 
			+ \mu_1^4\mu_5 
			+ \mu_2^2\mu_5}{(\mu_1^2 - \mu_2)(\mu_2^3 + \mu_3^2 + \mu_1^2\mu_4 - \mu_2(2\mu_1\mu_3 + \mu_4))} \\
		&&+ \frac{\mu_1^2 (3\mu_2^2\mu_3 + 2\mu_3\mu_4 - 2\mu_2\mu_5)}{(\mu_1^2 - \mu_2)(\mu_2^3 + \mu_3^2 + \mu_1^2\mu_4 - \mu_2(2\mu_1\mu_3 + \mu_4))},
\\
	&&\cdots\\ \nonumber
	b_{1}&=&
	-\mu_1^2 + \mu_2,\\
	b_{2}&=&
	-\frac{\mu_2^3 + \mu_3^2 + \mu_1^2\mu_4 - \mu_2(2\mu_1\mu_3 + \mu_4)}{(\mu_1^2 - \mu_2)^2},\label{a6}\\
	&&\cdots \nonumber
\end{eqnarray}
Replace the moments $\mu_n$ with $\tilde\mu_n$ on the right-hand side of these coefficients, we will obtain the Lanczos coefficients $\tilde a_n$ and $\tilde b_n$ for $\tilde f^{ W}(\omega)$. From Eqs. \eqref{a2}-\eqref{a6}, we find that
\begin{equation}
	\tilde a_{1}=-a_{1},\qquad \tilde a_{2}=-a_{2},\qquad \tilde b_{1}=b_{1},\qquad \tilde b_{2}=b_{2}.
\end{equation}
Similar relationships also exist for larger $n$, but due to the complexity of the expressions, we will not shown them in this work. In summary, we find that
\begin{equation}
	\tilde a_{n}=-a_{n},\qquad \tilde b_{n}=b_{n}.
\end{equation}

\section{Asymmetric truncation of a symmetric power spectrum}\label{appb}
It is observed that the symmetric truncation of a symmetric power spectrum will cause the Lanczos coefficients to eventually reach a plateau, whereas truncation on only one side will result in a single inflection point in the Lanczos coefficients. It naturally occurs to us that if the symmetric power spectrum is asymmetrically truncated on both sides, the Lanczos coefficients will exhibit two inflection points and eventually reach a plateau.

\begin{figure}[htb]
	\centering
	\subfigure[The variation of $\beta a_n$ with $n$.]{\includegraphics[width=0.45\textwidth]{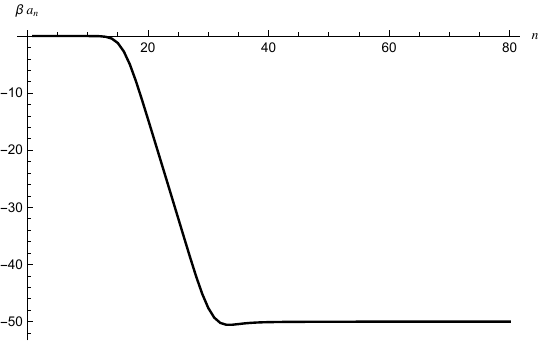}}
	\subfigure[The variation of $\beta b_n$ with $n$.]{\includegraphics[width=0.45\textwidth]{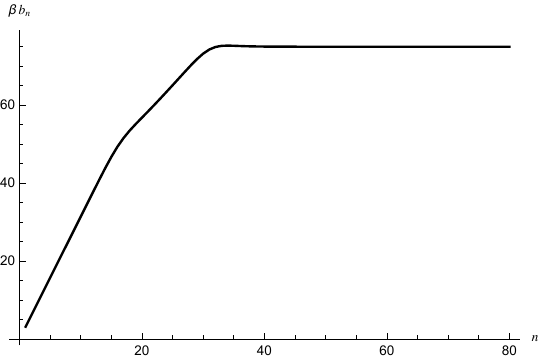}}
	\caption{Lanczos coefficients under asymmetric bilateral truncation of a symmetric power spectrum \eqref{b1} with $\beta=1$, truncated within $\omega\in[-200,100]$.}
	\label{abnthree}
\end{figure}

To investigate this issue, we consider the power spectrum
\begin{equation}\label{b1}
	f^{W}(\omega)=\mathcal{N}\frac{e^{\beta\omega/2}}{1+e^{\beta\omega}},\qquad -200\le \omega\le100.
\end{equation}
We find that the Lanczos coefficients indeed exhibit two inflection points and eventually reach a plateau, as shown in Figure \ref{abnthree}.

\FloatBarrier
\bibliographystyle{elsarticle-num}
\bibliography{references.bib}
\end{document}